\documentclass[aps,prl,floatfix,twocolumn,footinbib,amsmath,amssymb]{revtex4-1}
\usepackage{graphics}
\usepackage{epsfig}
\usepackage{natbib}
\usepackage{xspace}

\newcommand{\LiRb}{$^{6}$Li-$^{87}$Rb\xspace}
\newcommand{\ket}[1]{| #1 \rangle}
\newcommand{\bra}[1]{\langle #1 |}
\newcommand{\braket}[2]{\langle #1 | #2 \rangle}

\renewcommand{\paragraph}[1]{}
\newcommand{\citechin}{\cite{[{See }][{ and references therein.}]cold:chin10}\xspace}

\def\oper#1{\hat{\rm #1}}

\begin{document}

\title{Feshbach resonances of harmonically trapped atoms}
\author{Philipp-Immanuel Schneider, Yulian V. Vanne, and Alejandro Saenz}

       \affiliation{AG Moderne Optik, Institut f\"ur Physik,
         Humboldt-Universit\"at zu Berlin, Newtonstrasse 15,
         12489 Berlin, Germany}

       \date{\today}
 
\begin{abstract}
 Employing a short-range two-channel description we derive an analytic model 
 of atoms in isotropic and anisotropic harmonic traps at a Feshbach resonance. 
 On this basis we obtain a new parameterization of the energy-dependent 
 scattering length which differs from the one previously employed.
 We validate the model by comparison to full numerical calculations for \LiRb and
 explain quantitatively the experimental observation of a resonance shift and trap-induced molecules 
 in exited bands. Finally, we analyze the bound state admixture and
 Landau-Zener transition probabilities.
\end{abstract}

\maketitle

 In the last decade reams of fascinating experiments with ultracold atoms have 
 been carried out with applications ranging from studying condensed matter Hamiltonians
 and new phases of matter to performing quantum information processing \cite{cold:jaks05}. 
 Two key techniques made these achievements possible: (i) Atom-atom interaction characterized 
 by the $s$-wave scattering length $a$ can be tuned using a magnetic Feshbach 
 resonance (MFR). (ii) Atoms can be confined in various geometries such as
 dipole traps, optical lattices, or atomic waveguides \cite{cold:bloc08}.
 The known theory of MFR's successfully describes the {\em free} scattering process 
 for varying magnetic field $B$ and energy $E\rightarrow 0$. However, for the full
 understanding and precise controllability of {\em confined} atoms at an MFR a trap-version of 
 this theory is needed which incorporates the energy dependence of the scattering process.
 This is especially the case for strong confinement which has been lately used to
 explore confinement-induced resonances and scattering in mixed dimensions
 \cite{cold:hall10b,*cold:lamp10}.
 These systems show exciting behavior such as the formation
 of confinement-induced molecules \cite{cold:mori05}.

 In the following we present an approach to analytically describe MFR's of harmonically trapped atoms. 
 We obtain the eigenenergy equation, the admixture of the resonant molecular 
 bound state $A(E,B)$, and a general parameterization of the energy-dependent scattering length $a(E,B)$.
 We show that the energy dependence of $a$ differs significantly from the one previously used to 
 describe trapped gases \cite{cold:bold02,*cold:wout03,*cold:kryc09,cold:idzi06} while it
 confirms the functional form of other two-channel models for trapped atoms 
 \cite{cold:dunn05, *cold:moor06, *cold:nyga08}.
 We derive energy-dependent formulations of the resonance width and resonance
 shift and extend the model to anisotropic harmonic traps.
 The validity of our approach is verified by comparing to full multi-channel 
 calculations for \LiRb in harmonic confinement.

 We demonstrate the usefulness of the new model by explaining the experimental observation 
 of a shift of the resonance position of $^{87}$Rb in an optical lattice \cite{cold:wide04}
 and by circumstantiating the observation of confinement-induced molecules in exited states 
 by \textcite{cold:syas07}. We analyze the bound state admixture and show that it might be
 responsible for enhanced losses of trapped $^6$Li far away from the resonance \cite{cold:bour03a}. 
 Finally, Landau-Zener transition probabilities are derived from the full energy spectrum.

\paragraph{Hamiltonian}

 We consider the relative-motion $s$-wave Hamiltonian of two atoms in harmonic confinement
$\hat H = -\frac{\hbar^2}{2 m}\frac{d^2}{d R^2} + \frac12 m \omega^2 R^2 + \hat V_{ZH}
         + \hat V_{\rm int}(R)$ 
 with reduced mass $m$, trap frequency $\omega$, Zeeman and hyperfine energy $\hat V_{ZH}$, 
 and the electron-spin dependent interaction potential $\hat V_{\rm int}(R)$.
 Within the two-channel (TC) description one projects  
 onto the subspace of open and closed channels with the operators $\oper P$ and $\oper Q$,
 respectively. We consider the case of an elastic collision with one open channel.
 This results in the coupled equations
 \begin{eqnarray}
 \label{eq:two_channel_1}
  (\oper H_P -E) \ket{\Psi_P} + \oper W \ket{\Psi_Q} &=& 0 \,,\\
 \label{eq:two_channel_2}
  (\oper H_Q -E) \ket{\Psi_Q} + \oper W^\dagger \ket{\Psi_P} &=& 0\,, 
 \end{eqnarray}
 with $\oper H_P = \oper P \oper H \oper P$, $\oper H_Q = \oper Q \oper H \oper Q$,
 $\oper W = \oper P \oper H \oper Q$, $\ket{\Psi_P}=\oper P \ket{\Psi}$,
 $\ket{\Psi_Q}=\oper Q \ket{\Psi}$, and $E$ the energy above the threshold of the
 open-channel interaction potential \cite{cold:schn09a}.
 Furthermore, one assumes that close to the MFR $\ket{\Psi_Q}$ is simply a multiple $A$ 
 of a bound eigenstate  $\ket{\Phi_b}$ with eigenenergy $E_b$. We call this closed-channel
 state ``resonant bound state'' (RBS). To first order, the energy $E_b$ may be expanded 
 linearly in the magnetic field $B$, i.e. $E_b(B) = \mu (B-B_0)$.

 Be $\ket{\Phi_E}$ the normalized solution of the open channel with
 $\ket{\Psi_P} = C \ket{\Phi_E}$ then $C^2 + A^2=1$ holds
 which allows us to define a phase $\tan \delta_{\rm \scriptscriptstyle RBS}=A/C$ 
 attributed to the RBS admixture.
 Introducing $\ket{\Psi_P} = C \ket{\Phi_E}$ and 
 $\ket{\Psi_Q} = A \ket{\Phi_b}$ into Eq.~(\ref{eq:two_channel_2})
 and multiplying by $\bra{\Phi_b}$ gives 
\begin{equation}
 \label{eq:tan_res1}
 \tan{\delta_{\rm \scriptscriptstyle RBS}} = \frac{\bra{\Phi_E}\hat W\ket{\Phi_b}}{(E-E_b)}\,.
\end{equation}

 For infinite detuning $E-E_b$ the open channel is assumed to be in some
 background eigenstate $\ket{\Phi_{\rm bg}}$ of $\oper H_P$ with eigenenergy $E_{\rm bg}$.
 Multiplying Eq.~(\ref{eq:two_channel_1}) by $\ket{\Phi_{\rm bg}}$ and using 
 Eq.~(\ref{eq:tan_res1}) yields the eigenenergy equation
\begin{equation}
 \label{eq:TC_E-E_b}
  (E-E_b)(E-E_{\rm bg}) =  
    \frac{\bra{\Phi_{\rm bg}}\hat W\ket{\Phi_b} \bra{\Phi_E}\hat W\ket{\Phi_b}}
         { \braket{\Phi_{\rm bg}}{\Phi_E}}\,.
 \end{equation}
\paragraph{Short-range approximation}
 In order to find simplified expressions for $\bra{\Phi_{\rm bg}}\hat W\ket{\Phi_b}$, 
 $\bra{\Phi_E}\hat W\ket{\Phi_b}$, and $\braket{\Phi_{\rm bg}}{\Phi_E}$ we assume that
 the interaction acts only in some small range $R<R_{\rm int}\ll a_{\rm ho}$ such that 
 for $R>R_{\rm int}$ the solution $\ket{\Phi_E}$ is given by $\tilde\Phi_E(R) = 
 A_\nu D_\nu(\rho)$, where $D_\nu(\rho)$ is the parabolic cylinder function,
 $\rho=\sqrt{2} R/a_{\rm ho}$, $a_{\rm ho} = \sqrt{\hbar/(m\omega)}$,
 $\nu=E/(\hbar\omega)-1/2$, and $A_\nu$ is a normalization constant.
 For $R\ll a_{\rm ho}$ one has $D_\nu(\rho)=D_\nu(0)\left(1-\rho/[\sqrt{2}f(E)]\right)$ with
 $f(E) = \Gamma\left(\frac14-\frac{E}{2\hbar\omega}\right)/2\Gamma\left(\frac34-\frac{E}{2\hbar\omega}\right)$
 \cite{cold:abra65}. Considering the logarithmic derivative one obtains the scattering length
 $a(E) = a_{\rm ho} f(E)$, which is equivalent to the result in \cite{cold:busc98}.

 In the spirit of a Taylor expansion we parameterize $\bra{\Phi_E}\hat W\ket{\Phi_b}$ by a linear combination
 $\alpha\, \tilde \Phi_E(0) + \beta\, \tilde \Phi_E'(0)$. That is, one can define a 
 $\gamma_E$ and an $a^*$ such that
 \begin{equation}
 \label{eq:gamma_E}
  \bra{\Phi_E}\hat W\ket{\Phi_b} = \gamma_E \tilde \Phi_E(0)\left(1-\frac{a^*}{a}\right)\,.
 \end{equation}
 Here, $\gamma_E$ describes the coupling strength to the RBS and $a^*$ defines the scattering 
 length of the state $\ket{\Phi_E}$ when it is orthogonal to $\hat W\ket{\Phi_b}$.
 Since the orthogonality fixes the phase of $\ket{\Phi_E}$ within the coupling range
 the energy-dependence of $a^*$ should be usually negligible.
 We find that also the variation of $\gamma_E$ is negligible which can be explained by 
 the stability of the nodal structure of $\ket{\Phi_E}$ for most of the coupling range.
 Analogous to Eq.~(\ref{eq:gamma_E}), we set $\bra{\Phi_{\rm bg}}\hat W\ket{\Phi_b} = 
 \gamma_{\rm bg} \tilde \Phi_{E_{\rm bg}}(0)\left(1-\frac{a^*}{a_{\rm bg}}\right)$
 where we allow for a different coupling strength $\gamma_{\rm bg}$ of the uncoupled 
 background state to the RBS. We assume $a^*$ to have the same value as in Eq.~(\ref{eq:gamma_E})
 since it is determined by the requirement of orthogonality to the constant term $\hat W\ket{\Phi_b}$.

 Finally, we set $\braket{\Phi_{\rm bg}}{\Phi_E}=\braket{\tilde \Phi_{\rm bg}}{\tilde \Phi_E}$ neglecting
 the behavior of the wave-functions at $R<R_{\rm int}$. 
 This approximation cannot reproduce the exact energies $E$ where 
 $\braket{\Phi_{\rm bg}}{\Phi_E}=0$ that depend on the nodal structure at $R<R_{\rm int}$.
 However, it is applicable in a sufficient range around $E=E_{\rm bg}$  such that states of
 any energy can be described by choosing an appropriate background state.

 Applying these assumptions to Eq.~(\ref{eq:TC_E-E_b}) 
 and using the properties of $\braket{D_\nu}{D_{\nu'}}$ \cite{cold:grad07} one finds
 the analytic eigenenergy equation
 \begin{equation}
 \label{eq:eigenenergies2}
       E-E_b = \frac{2 \gamma_{\rm bg}\gamma_E}{a_{\rm ho} \hbar\omega} 
                 \frac{\left(f(E)-\frac{a^*}{a_{\rm ho}}\right)\left(f(E_{\rm bg})-\frac{a^*}{a_{\rm ho}}\right)}
                      {f(E)-f(E_{\rm bg})}\,.
 \end{equation}

\paragraph{Energy-dependent scattering length}
 In order to determine the interaction dependent scattering length $a(E,B)$ we demand 
 that it is equal to the scattering length $a_{\rm ho}f(E)$ of $\tilde \Phi_E(R)$, i.e.\
 that the eigenenergies are given by the roots of $a(E,B)=a_{\rm ho}f(E)$.
 Since $R_{\rm int}\ll a_{\rm ho}$ the trap has no influence 
 on the interaction and thus on $a(E,B)$.
 The value of $f(E_{\rm bg})$ in Eq.~(\ref{eq:eigenenergies2}) is in analogy determined by the 
 root of the eigenequation for the uncoupled problem $a_{\rm bg}(E_{\rm bg})=a_{\rm ho}f(E_{\rm bg})$
 which is closest to $E$. 
 Here, $a_{\rm bg}$ is the background scattering length that varies with the energy approximately like
 $a_{\rm bg}(k)^{-1} = a_0^{-1} - \frac12 k^2 R_{\rm eff}$ with $k^2=2m E/\hbar^2$, $a_0$ the 
 zero-energy background scattering length, and $R_{\rm eff}>0$ the effective range that can be well 
 estimated from the van-der-Waals coefficient $C_6$ \cite{cold:flam99}.
 Since the interaction is trap independent one can use the $\omega \rightarrow 0 $ limit to set 
 $f(E_{\rm bg}) \equiv a_{\rm bg}(E)/a_{\rm ho}$.
 Then, rearranging Eq.~(\ref{eq:eigenenergies2}) yields
\begin{equation}
\label{eq:a_E_trap}
 a(E,B) = a_{\rm bg}(E)\left(1-\frac{\Delta B}{B - B_0 + \delta B - E/\mu}\right) 
\end{equation}
 with resonance width 
 $\Delta B= \frac{2 \gamma_{\rm bg}\gamma_E m a_{\rm bg}}{\mu \hbar^2}\left(1-\frac{a^*}{a_{\rm bg}}\right)^2$ 
 and detuning $\delta B = a_{\rm bg} \Delta B/(a_{\rm bg}-a^*)$. 
 The scattering length $a^*=a(E=0,B_0)$ has an important impact on the behavior of $\Delta B$ and $\delta B$.
 For small $a_{\rm bg}/a^*$ one has $\Delta B \propto a_{\rm bg}^{-1}$ and $\delta B \approx$ const.\ 
 while for systems with large $a_{\rm bg}$ the value of $a^*$ is negligible such that 
 $\delta B = \Delta B \propto a_{\rm bg}$.

 Let us compare the result to the previously used energy dependence of the scattering length 
 given as \cite{cold:bold02,*cold:wout03,*cold:kryc09,cold:idzi06} 
\begin{equation}
\label{eq:freecase2}
 a(E,B) = a_{\rm bg}\left(1-\frac{\Delta B\left(1+(k a_{\rm bg})^2\right)}
                                         {B - B_0 + \delta B + (k a_{\rm bg})^2\Delta B - E/\mu}
                            \right)\,. 
\end{equation}
 Here, the term $(k a_{\rm bg})^2$ induces an additional energy-dependence.
 We examined a two-channel model system and 
 found no energy dependence connected to $(k a_{\rm bg})^2$ while the behavior described by 
 Eq.~(\ref{eq:a_E_trap}) could be validated \footnote{A detailed analysis will be published elsewhere.}.
 The absence of the dependence on $(k a_{\rm bg})^2$ is also supported by other two-channel models in the 
 presence of a trapping potential \cite{cold:dunn05, *cold:moor06, *cold:nyga08}.

\paragraph{RBS admixture}
 
 From the formal limit $E_{\rm bg}\rightarrow E$ of Eq.~(\ref{eq:eigenenergies2}) and of the 
 short-range approximation of Eq.~(\ref{eq:TC_E-E_b}) one can infer that 
 $\tilde\Phi_E^2(0) = \frac{2}{a_{\rm ho}\hbar\omega}\frac{f^2(E)}{f'(E)}$. With this the short-range
 approximation of Eq.~(\ref{eq:tan_res1}) can be written in terms of $f(E)$ as
\begin{equation}
 \label{eq:tan_res_SR}
 \tan^2 \delta_{\rm \scriptscriptstyle RBS} 
    = \frac{\gamma_E}{\gamma_{\rm bg}} 
      \frac{a_{\rm ho}}{a_{\rm bg} \mu \Delta B}
      \frac{\left(f(E) - a_{\rm bg}/a_{\rm ho}\right)^2}{f'(E)}\,.
\end{equation}

\paragraph{Extension to anisotropic harmonic traps}

 Our model can be easily extended to anisotropic harmonic traps.
 The eigenenergy relation in a trap with $\omega_x=\omega_y=\eta \omega_z$ is known to be 
 $a = -\sqrt{\pi} d /\mathcal F(x,\eta)$ with $d,x,$ and $\mathcal F$ defined in \cite{cold:idzi06}. 
 Since $a(E,B)$ is trap-independent we have to replace for $\eta\neq1$  
 in the eigenenergy relation $a_{\rm ho}f(E)$  by $-\sqrt{\pi} d /\mathcal F(x,\eta)$. 
 One can show that this necessitates the same replacement in the
 expression for $\tilde \Phi_E^2(0)$ and accordingly in Eq.~(\ref{eq:tan_res_SR}).

\paragraph{Comparison with full numerical calculations}
 For the realistic case of \LiRb we have performed full numerical multi-channel (MC) calculations in 
 order to obtain eigenenergies and channel admixtures for different trap frequencies $\omega$ ($\eta=1$)
 and magnetic fields $B$.
 From the $E\rightarrow 0$ limit in free space we obtain $a_0 = -17.77\,$a.u. 
 and from the $C_6$ coefficient $R_{\rm eff} = 1899.9\,$a.u.
 The magnetic field positions of vanishing and resonant scattering length
 and the channel admixtures at resonance
 of the first two trap states in a shallow trap with $\omega = 2\pi \times 20\,$kHz yield the
 parameters $\mu = 2.44\,\mu_B$, $a^* = 63.14\,$a.u., $B_0 = 1064.62\,$G,
 $\gamma_E = 2.73\times 10^{-8}\,$a.u., and $\gamma_{\rm bg} = 2.63\times 10^{-8}\,$a.u.
 
\begin{figure}[htp]
 \centering
 \includegraphics[width=0.45\textwidth]{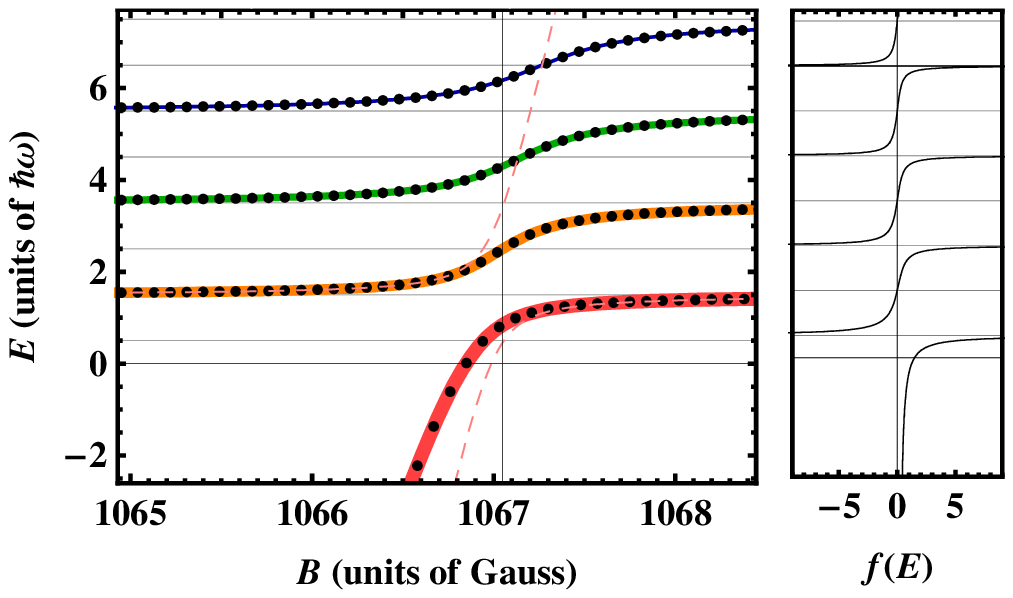}
 \includegraphics[width=0.44\textwidth]{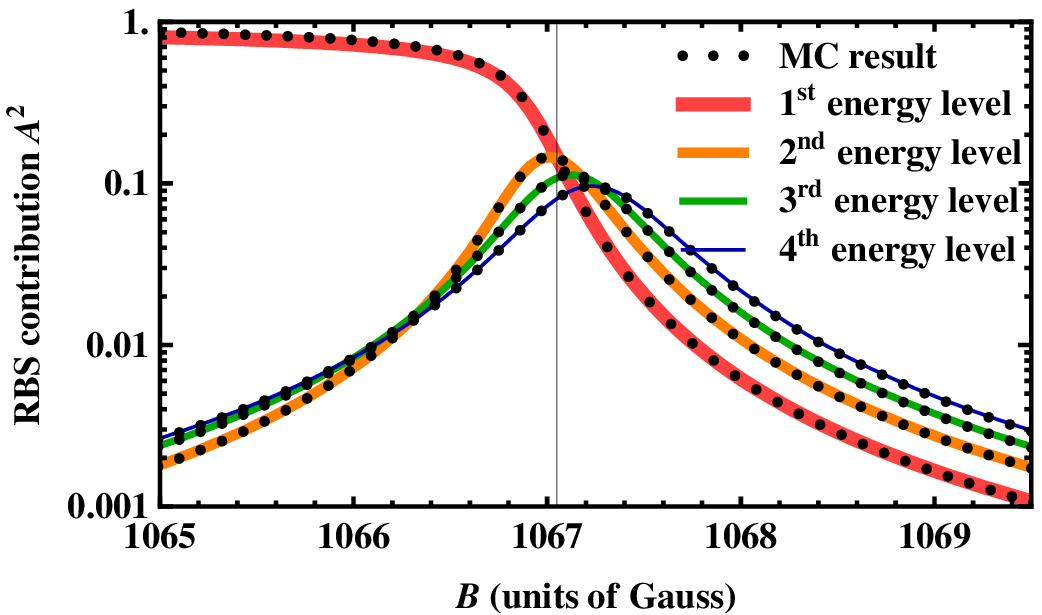}\\
 \vspace{-9.25cm}
 \hspace{-0.1cm} a) \hspace{4.6cm} b)\\
 \vspace{4.5cm} \hspace{-4.9cm} c)
 \vspace{4cm}
 \caption{
 a) Energy spectrum of \LiRb as a function of the magnetic field $B$ 
 in a trap with $\omega=2\pi\times200\,$kHz. Dots indicate MC calculations while lines indicate
 solutions of Eq.~(\ref{eq:eigenenergies2}). 
 The eigenenergies of Eq.~(\ref{eq:avoided_crossing}) are shown as dashed lines. 
 b) Scattering length $a(E,B)/a_{\rm ho}=f(E)$ corresponding to an eigenenergy $E$ of the system. 
 At $E=\hbar\omega(2n+1/2)$ with $n=\mathbb N$ resonances appear. 
 c) Contribution $A^2$ of the RBS for each energy level as a function of the magnetic field $B$.
    Dots indicate MC calculations while lines indicate results of Eq.~(\ref{eq:tan_res_SR})
    with energies from solutions of Eq.~(\ref{eq:eigenenergies2}). }
 \label{fig:spectrum}
\end{figure}

 Figure~\ref{fig:spectrum} shows a comparison of the eigenenergies and RBS admixture 
 obtained from the full MC calculation and from our model in a trap with
 $\omega = 2\pi \times 200\,$kHz which corresponds to a deep optical lattice.
 Both results are in very good agreement with a deviation $<0.003\,\hbar\omega$
 and $<0.1\%$, respectively. This shows that the model accurately covers the $E$
 and $B$ dependence of the scattering process.
 Only for energies well below zero, the model fails to reproduce $E$ and $A$ correctly.
 Here, the van-der-Waals interaction becomes dominant the long range such that the 
 approximation of a short-rang interaction breaks down.
 For the considered trap the roles of $a^*$ and $a_{\rm bg}(E)$ become apparent through a 
 significant broadening of $\Delta B$ by $0.18\,$G between the $1^{\rm st}$ and the $4^{\rm th}$ state. 

 \paragraph{Resonance position in the harmonic trap}
 As shown in Fig.~\ref{fig:spectrum} b) the resonances of the scattering length $a(E,B) \propto f(E)$ 
 are located at $E_{\rm res}^{(n)}=\hbar\omega(2n + \frac12)$. In an anisotropic trap the resonance
 energies $E_{\rm res}^{(n)}$ are determined by the roots of $\mathcal F(x,\eta)$.
 Consequently, the magnetic resonance position changes according to Eq.~(\ref{eq:a_E_trap}) from the 
 free-space position $B_R = B_0 - \delta B$ to
\begin{equation}
 \label{eq:B_res_n}
 B_{\rm res}^{(n)} = B_0 - \delta B + E_{\rm res}^{(n)}/\mu\,.
\end{equation}
 The difference of the resonance position for each energy level $n$ opens the exciting possibility 
 to tune the magnetic field to a resonance of a specific trap state which in turn enhances 
 inelastic collisions depopulating this level. 
 By successively adjusting the magnetic field at different resonance positions one might 
 be able to engineer an ensemble in an excited state or cool the system to its 
 relative-motion ground state. 
 A good candidate for this approach would be an MFR of $^{133}$Cs at $19.8\,$G
 where the small value of $\mu \Delta B = 2\pi\times 4\,$kHz \citechin 
 admits to address single levels in reasonably deep traps.

 Applying Eq.~(\ref{eq:B_res_n}) one is able to explain the disagreement of an 
 experimentally observed MFR position of $^{87}$Rb in a negligibly weak trap ($B_{\rm res} = 9.09(1)\,$G 
 \cite{cold:erha04}) and a trap of frequency $\omega_x = 2\pi\times 33\,$kHz, 
 $\omega_y\approx\omega_z\approx 2\pi\times42\,$kHz ($B_{\rm res} = 9.121(9)\,$G 
 \cite{cold:wide04}).
 The energy dependence of $\delta B$ in unknown. However, its impact is likely to be negligible.
 For $C_6 = 4660\,$a.u. \cite{cold:klau01} it holds $\left. d\,\delta B/d E\right|_{E=0} < 0.1\, 
 \mu_B^{-1}$ for both $a_{\rm bg}\ll a^*$ and $a_{\rm bg}\gg a^*$ while $1/\mu = 0.5\, 
 \mu_B^{-1}$ \citechin.
 Hence, the resonance shift is approximately given by $E_{\rm res}^{(1)}/\mu = 0.034\,$G 
 which is in good agreement with the experimental results.

\begin{figure}[htp]
 \centering
 \includegraphics[width=0.45\textwidth]{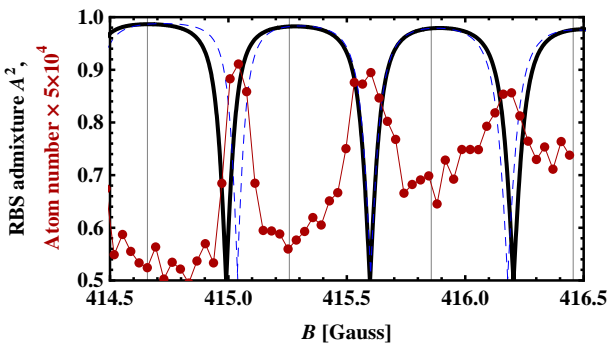}
 \caption{
 Atom number vs. magnetic field (dots) as measured in \cite{cold:syas07}. 
 RBS admixture of eigenstates according to Eq.~(\ref{eq:tan_res_SR}) with 
 $\gamma_E=\gamma_{\rm bg}$, $a^* \gg a_{\rm bg}$, and $a_{\rm bg}=a_0$ (solid line) 
 and for an energy-dependent $a_{\rm bg}$ given in the text (dashed line).}
 \label{fig:rempe}
\end{figure}

\paragraph{Trap-induced molecules in exited states}
 Another effect of the trap concerns the RBS admixture.
 It is present for each energy level [see Fig.~\ref{fig:spectrum} c)] which suggests that
 RBS molecules can be created not only in the bound state \cite{cold:part05}
 but also in exited states, e.g.\ exited bands of an optical lattice. 
 Indeed, these confinement-induced molecules have been experimentally 
 observed by \textcite{cold:syas07}.
 By inducing Rabi oscillations between atoms and RBS molecules at a very narrow $^{87}$Rb 
 resonance ($\mu \Delta B = 2\pi\times 2\,$kHz, $a_{\rm bg}=100.8\,$a.u.) 
 they could produce a large number of molecules in an optical lattice with two atoms per 
 site in the center. After a sudden change of the magnetic field they measured the number
 of unbound atoms featuring pronounced maxima and broad minima.
 The suppressed dissociation at the minima can be attributed to a strong RBS admixture of 
 excited trap states.
 Supported by our MC calculations for \LiRb we assume $\gamma_E\approx\gamma_{\rm bg}$ in order
 to estimate the RBS admixture using Eq.~(\ref{eq:tan_res_SR}). 
 Figure~\ref{fig:rempe} shows the atom number observed in \cite{cold:syas07} and the RBS admixture 
 for excited eigenstates at different magnetic fields. Clearly, a large RBS admixture coincides 
 with suppressed dissociation. Here, the RBS molecules survive as part of an eigenstate of the 
 new Hamiltonian while for magnetic fields where the RBS admixtures are small the projection of the
 RBS onto the eigenstates leads to a larger fraction of unbound atoms.
 The effects of the energy-dependence of $a_{\rm bg}$ can be studied by an effective range 
 approximation $a_{\rm bg} = (a_0^{-1} - \frac12 k^2 R_{\rm eff}
 + \frac14 k^4 V_{\rm eff})^{-1}$. With $a_0=100.8\,$a.u., $R_{\rm eff}$ determined from 
 $C_6 = 4660\,$a.u. \cite{cold:klau01} and $V_{\rm eff} = (1700\,{\rm a.u.})^3$ the
 positions of small RBS admixture are shifted towards those of maximal 
 dissociation.

\paragraph{Novel resonance phenomenon}
 For no external trap the resonance of the scattering length coincides with the maximal 
 RBS admixture to the scattering wave function \cite{cold:schn09a} such that the influence 
 of both effects can hardly be distinguished. In traps this rule can be strongly violated. 
 Searching for roots of the derivative of Eq.~(\ref{eq:tan_res_SR}) with respect to $E$ one finds 
 a maximal RBS admixture where $E$ solves $a_{\rm bg}(E)/a_{\rm ho} = f(E) - 2 f'(E)^2/f''(E)$.
 For large $a_{\rm bg}$ and higher lying solutions $E_{\rm max}^{(n)}$ of this equation
 $E_{\rm max}^{(n)}$ can be far away from the resonance energies $E_{\rm res}^{(n)}$.
 Translated to the magnetic field the offset
 between the resonance position $B_{\rm res}^{(n)}$ and the position of maximal RBS 
 admixture $B_{\rm max}^{(n)}$ can even approach $\Delta B$ which is accompanied by
 a vanishing scattering length at $B_{\rm max}^{(n)}$!
 Hence, this offset should be significant in a Fermionic system such as $^6$Li with 
 a large background scattering length.

 \textcite{cold:bour03a} performed an experiment with $2N=7\times10^4$ $^6$Li atoms in two different 
 hyperfine states in a trap with  $\omega_x=2\pi\times 0.78\,$kHz and $\omega_y\approx 
 \omega_z\approx 2\pi\times 2.2\,$kHz \cite{cold:bour03a}.
 They found a \emph{local} maximum of atom loss 
 close to $B_R$ but a \emph{global} one at an about $-80\,$G shifted magnetic field.
 Atoms at the Fermi edge have a relative-motion energy equal to the Fermi energy 
 $E_F= 2\pi\hbar (6N \omega_x \omega_y \omega_z)^{1/3}$. For $a_0=-1405\,$a.u., 
 $\Delta B|_{E=0}=-300\,$G, $C_6 = 1393.4\,$a.u.\ \citechin and $\eta=3$ our model
 predicts a maximal RBS admixture $-80.8\,$G shifted from the resonance. 
 This agrees well with the maximum loss position which can be an indication that the RBS admixture
 enhances transitions to deeper bound states and thereby influences atom-loss processes. Note, that 
 another qualitative explanation for the off-resonant loss has been given by 
 \textcite{cold:bour03a}.

\paragraph{Landau-Zener avoided crossing}
 Finally, we derive Landau-Zener transition propabilities for each avoided
 crossing in the spectrum.
 Expanding $f(E) \approx f(E_{\rm bg}) + f'(E_{\rm bg})(E-E_{\rm bg})$ in 
 Eq.~(\ref{eq:eigenenergies2}) around some background energy $E_{\rm bg}$ yields
 the eigenenergy equation
\begin{equation}
\label{eq:avoided_crossing}
 (E-E_b-\mu\delta B)(E-E_{\rm bg}) = \mu \Delta B \frac{f(E_{\rm bg})}{f'(E_{\rm bg})}
\end{equation}
 which describes the avoided crossing of a molecular eigenstate to a background state
 with coupling strength $\delta^2 = \mu \Delta B f(E_{\rm bg})/f'(E_{\rm bg})$.
 For the $n^{\rm th}$ avoided crossing and $a_{\rm bg}\ll a_{\rm ho}$ we have
 $1/f'(E_{\rm bg})\approx 2 \hbar\omega \binom{n-1/2}{n-1}/\sqrt{\pi}$
 \cite{cold:busc98} such that the diabatic transition probability is given as $P_n=e^{-2\pi G_n}$
 with $ G_n = |\delta^2/(\hbar \mu \dot B)| \approx
  \left| 2 \omega \Delta B a_{\rm bg} \binom{n-1/2}{n-1}/ (\sqrt{\pi} a_{\rm ho} \dot B )\right|$.
 Of course, only for $|\delta|\ll\hbar\omega$ the Landau-Zener theory can give exact 
 results while otherwise two coupled states offer only a quantitative
 approximation. This can be judged from Fig.~\ref{fig:spectrum} a)
 for the first avoided crossing where $|\delta|\approx1.4\hbar\omega$.

 In conclusion, we developed an analytic model of atoms in isotropic and anisotropic harmonic 
 traps experiencing a Feshbach resonance. The energy-dependent scattering length was determined
 and compared to a previous parameterization.
 Consequences of the model including a resonance shift, molecules in excited trap states, and
 a maximal molecular admixture away from the resonance were studied.
 Our model is in agreement with full numerical calculations and experimental results.
 We expect the approach to be applicable for an analytic treatment of other Feshbach-type 
 resonances in a quasicontinuum. 

 \acknowledgments{We are grateful to the {\it Deutsche Forschungsgemeinschaft} (SFB\,450),
 the {\it Fonds der Chemischen Industrie}, and the {\it Deutsche Telekom Stiftung} for financial support.}
%
%

\end{document}